\documentclass[preprint,12pt]{elsarticle}
\usepackage{amsmath}
\usepackage{amssymb,amsfonts,textcomp}
\usepackage{color}
\usepackage{array}
\usepackage{supertabular}
\usepackage{hhline}
\usepackage{hyperref}

\makeatletter
\newcommand\arraybslash{\let\\\@arraycr}

\begin{document}

\begin{frontmatter}

\title{
Effects of atoms and molecules adsorption on electronic and magnetic properties of s-triazine with embedded Fe
atom: DFT investigations}


\author[a,b]{Yusuf Zuntu Abdullahi\corref{cor1}}
\ead{yusufzuntu@gmail.com}

\cortext[cor1]{Corresponding author}
 
\author[a]{Tiem Leong Yoon}
\ead{tlyoon@usm.my} 

\author[a]{Mohd Mahadi Halim}
\author[a]{Md. Roslan Hashim}
\author[c]{Thong Leng Lim}

\address[a]{School of Physics, Universiti Sains Malaysia, 11800 Penang, Malaysia}
\address[b]{Department of Physics, Faculty of Science, Kaduna State University, P.M.B. 2339 Kaduna State, Nigeria}
\address[c]{Faculty of Engineering and Technology, Multimedia University, Jalan Ayer Keroh Lama, 75450 Melaka, Malaysia}







\begin{abstract}
We employ first-principles calculations to study the mechanical, geometrical, electronic and magnetic properties of Fe atom embedded s-triazine ($\mathrm{Fe}$@${\mathrm C_6}{\mathrm N_6}$) system under the influence of external environment. Our results show that the binding energy of $\mathrm{Fe}$@${\mathrm C_6}{\mathrm N_6}$ can be modulated by an applied tensile deformation and perpendicular electric field. The non-magnetic semiconducting property of pure s-triazine sheet (${\mathrm C_6}{\mathrm N_6}$) is found to change upon embedding of Fe atom in the porous site of the sheet. It is revealed that the $\mathrm{Fe}$@${\mathrm C_6}{\mathrm N_6}$ system exhibits half-metallic electronic character with a magnetic moment in the the order similar to that of an isolated Fe atom. Furthermore, electronic and magnetic properties of the $\mathrm{Fe}$@${\mathrm C_6}{\mathrm N_6}$ systems are preserved up to a maximum value of 10 V/nm in electric field strength and 6\% tensile strain. Interestingly, we find that the half-metallic electronic character of $\mathrm{Fe}$@${\mathrm C_6}{\mathrm N_6}$ system can be tuned into a semiconductor via adsorption of atoms and molecules into the $\mathrm{Fe}$@${\mathrm C_6}{\mathrm N_6}$ system. The magnetic moment of $\mathrm{Fe}$@${\mathrm C_6}{\mathrm N_6}$ with adsorbed atoms/molecules is also modified. Our findings may serve as a guide for future applications of $\mathrm{Fe}$@${\mathrm C_6}{\mathrm N_6}$  structures in spintronics devices.

\end{abstract}
\end{frontmatter}

\section{Introduction}

Currently there is an intense search for suitable substrates for transition metal (TM) atoms encapsulation \cite{Abdullahi:CAP2016},\cite{Zhang:PLA2016},\cite{Ghosh:JMCC2014}.
The central concern in the encapsulation is to ensure that the substrate remains inert and strongly
binds to the TM atoms. Moreover, the appropriate substrate is expected to preserve its intrinsic properties and that of
bound TM atoms. Two-dimensional (2D) carbon-based and related surfaces with compacted hexagonal rings have been a frequent choice for trapping TM atoms \cite{AlZahrani:PBCM2012},\cite{Abdullahi:QM2015},\cite{Ersan:ASC2016},\cite{Rahman:JCTN2015}  
due to their desirable surface properties. Numerous works have so far
been done to investigate the stable geometries and electronic properties of TM atoms adsorption on graphene and boron nitride sheets \cite{Rahman:JCTN2015},\cite{Zhang:PELSN2015},\cite{Yagi:PRB2004}.
However, their reports have shown that the adatoms tight weekly on these 2D surfaces as a result
of low adsorption energies. Additionally, the large surface free energy of the TM atoms would make them aggregate
easily to form cluster on these surfaces. To ensure the immobility of the TM atoms on the surface, various defect
sites have been proposed \citep{Zhang:PELSN2015}.
Creation of defect sites would presumably bind the TM atoms firmly to the surface. However, forming regular defect sites experimentally may not be possible due to influence of external environment. Thus, much efforts have been made to synthesis 2D materials with inherently defined porous sites \cite{Kroke:NJC2002}.

Among the recently synthesized porous 2D materials \cite{Asadpour:SSC2015},
graphitic carbon nitride ($\mathrm{CN}$) sheet has received
much attention \cite{Wang:N2014}, \cite{Abdullahi:SSC2016}.
This is due to its potential as a right candidate for various applications \cite{Wang:NN2012},\cite{Li:CS2012},\cite{Zhang:CS2012},\cite{Zhang:JACS2012},\cite{Zhang:AM2014}.
Graphitic CN belongs to a group of allotrope with a common chemical formula  g-$\mathrm{C}_x\mathrm{N}_y$, where $x$ and $y$ represent the number of C and N atoms in the unit cell. For example, single triazine-based graphitic CN has a chemical formula g-s-${\mathrm C}_3{\mathrm N}_4$ whereas tri-single triazine-based (heptazine) graphitic CN is named g-t-s-$\mathrm{C}_3\mathrm{N}_4$ \cite{Abdullahi:SSC2016}.
The hexagonal rings in the unit cell of heptazine are compacted in an abreast manner. s-triazine with a chemical formula g-${\mathrm C_6}{\mathrm N_6}$ is another allotrope of graphitic CN in which two of its hexagonal rings (g-$\mathrm{C_3}\mathrm{N_3}$) are separated via a C-C bond \cite{Wang:N2014}.
Depending on the composition of C and N in the hexagonal rings and unit cell, these allotropes can have different
electronic properties ranging from semiconducting to half-metallic. For instance, triazine-based  g-$\mathrm{C_4N_3}$ is another allotrope of graphitic CN which possess a ferromagnetic ground state with half metallic electronic character \cite{Du:PRL2012}
whereas the rest of the allotropes are non-magnetic with wide or small band gaps \cite{Wang:N2014},\cite{Abdullahi:SSC2016}.

To search for a suitable spintronic material such as diluted magnetic semiconductor, first-principles calculations of TM
atoms embedded in semi-conducting CN sheet have been performed \cite{Abdullahi:CAP2016},\cite{Zhang:PLA2016},\cite{Ghosh:JMCC2014}. 
Choudhuri {\it et al.}. recently reported the findings
of TM embedded  g-$\mathrm{C_3N_3}$ (TM@g-$\mathrm{C_3N_3}$) systems based on density functional theory (DFT) \cite{Choudhuri:JMCC2016}.
The TM atoms (Cr, Mn
and Fe) embedded  g-$\mathrm{C_3N_3}$ systems are found to be dynamical, thermally and mechanically stable. Their report also show that Cr, Mn and Fe embedded in the g-t-C\textsubscript{3}N\textsubscript{3} systems exhibit high temperature
ferromagnetism and high magnetic anisotropy energy (MAE) with a peak value per atom occurring in
Cr@g-t-C\textsubscript{3}N\textsubscript{3}. 

TM embedded graphitic carbon nanostructures can also be used for heterogeneous catalysis, hydrogenation of
CO\textsubscript{2} and as membrane for separation of gases \cite{Sun:SR2013},\cite{Ji:RSCAdv2016},\cite{Ma:IJHE2014},\cite{Ma:C2015},\cite{Sirijaraensre:ASS2016}.
A recent theoretical study predicts excellent
catalytic activities for single atomic catalyst (SAC) of TM embedded C\textsubscript{2}N systems \cite{Ma:C2015}.
Additionally, an
effective catalysis of hydrogenation of CO\textsubscript{2} into methanol on the copper embedded graphene system has been  demonstrated by Sirijaraensre and Limtrakul \cite{Sirijaraensre:ASS2016}.
Although good catalytic behavior has been reported for the SAC of TM
embedded C\textsubscript{2}N systems, a comprehensive understanding of various types of atoms and molecules adsorbed
onto their surfaces is still lacking. In this work, we investigate the structural, electronic and magnetic
properties of Fe embedded s-triazine sheet with various adsorbed atoms and molecules. We begin by analyzing
the Fe embedded s-triazine without the adsorbed atoms and molecules. Then, we introduce these atoms (C,N,O,H) and molecules ($\mathrm{CH}_4,{\mathrm N_2},{\mathrm O_2},{\mathrm H_2},\mathrm{CO},\mathrm{CO}_2$) at a distant height above the porous site at which the Fe is situated. The relaxed Fe embedded s-triazine with adsorbed atoms and molecules systems are then further analysed.

\section{Conputational method}
We use QUANTUM ESPRESSO package \cite{Giannozzi:JPCM2009}
to carry out spin-polarised DFT \cite{Hohenberg:PR1964}
calculations within 
Perdew-Burke-Enzerhof (PBE) approximation \cite{Perdew:PRL1996}.
Ultrasoft pseudopotential method is employed to treat the core and valence electrons of C, N, and Fe (semi-core included) atoms \cite{Vanderbilt:PRB1990}.
A plain wave basis set with kinetic energy cut-off of 550 eV was used to expand the wave functions. Marzari-Vanderbilt smearing method with Gaussian spreading is employed \cite{Marzari:PRL1999}
to aid the convergences of our systems. The Brillouin zone (BZ) is sampled with $8 \times 8 \times 1$ and $15 \times 15 \times 1$ Mankhorst-Pack k-point meshes for the self-consistent field and density of
state calculations respectively \cite{Monkhorst:PRB1976}.
The planar sheet is treated with periodic boundary condition and a thick vacuum space of $16$ \AA{} along the $z$-direction to avoid interaction between periodic layers. All geometries are fully
relaxed using the BFGS quasi-Newton algorithm until Hellmann-Feynman force tolerance on each atom was smaller than 
0.05 $\mathrm{eV}$/\AA.

\begin{figure}
{\centering  \includegraphics[width=6.5in,height=2.5902in]{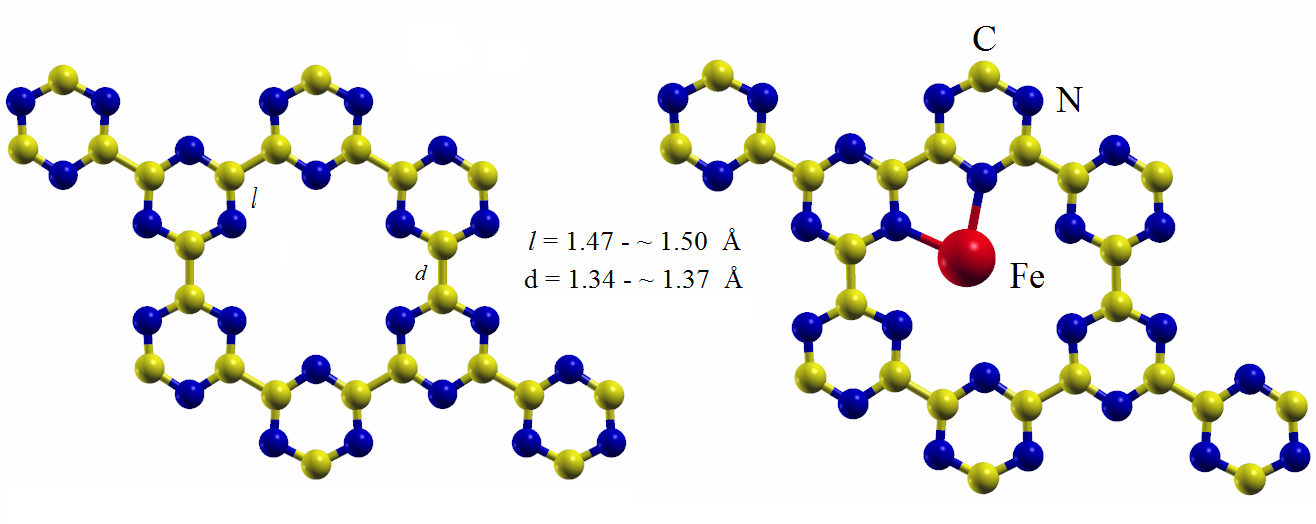} \par}
{\centering (a)\par}
{\centering  \includegraphics[width=4.0in,height=0.8in]{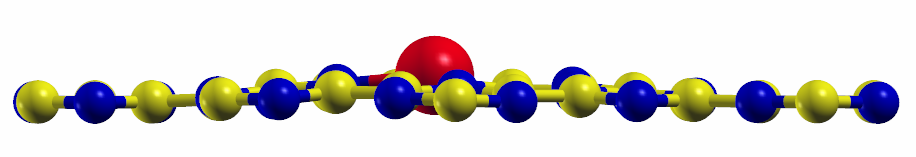} \par}
{\centering (b)\par}
{\centering  \includegraphics[width=3.6874in,height=3.502in]{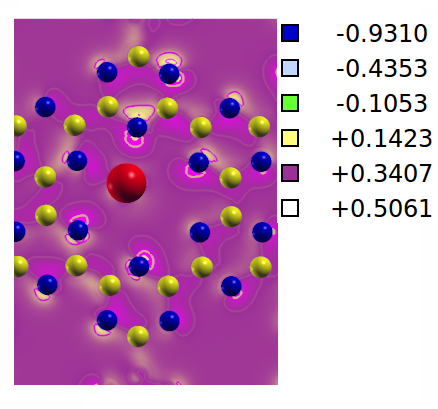} \par} {\centering (c) \par}
\caption{
{(a)}. Relaxed structure of 2{\texttimes}2 C\textsubscript{6}N\textsubscript{6} sheet (left
panel) and relaxed structure (right) of Fe embedded 2{\texttimes}2  ${\mathrm C_6}{\mathrm N_6}$. {(b). }Relaxed side view of 2{\texttimes}2  ${\mathrm C_6}{\mathrm N_6}$ sheet with an embedded Fe atom
(Fe@C\textsubscript{6}N\textsubscript{6}) under perpendicular electric field strength of 10 V/nm. {(c).} Difference charge-density for Fe atom embedded s-triazine. The color scale shows ranges of charge accumulation and depletion in a.u.}
\label{fig1}
\end{figure}

The optimized geometries of $2 \times 2$ pure s-triazine sheet (C\textsubscript{6}N\textsubscript{6}) and
Fe embedded C\textsubscript{6}N\textsubscript{6} sheet (Fe@C\textsubscript{6}N\textsubscript{6}) are displayed in Fig.~\ref{fig1}(a). All N atoms have sp\textsuperscript{3}-like hybridized structure, whereas each C atom which is bonded to three atoms has sp\textsuperscript{2}{}-like hybridized structure. The relaxed C-C bond length \textit{d} linking the
s-triazine units in  $\mathrm{Fe}$@${\mathrm C_6}{\mathrm N_6}$ system and C-N bond lengths \textit{l} in the embedded Fe atom
cavity are in the range of 1.47-1.50 \AA\textit{} and 1.35-1.37 \AA\textit{} respectively. The calculated values agree
well with the reported results \cite{Wang:N2014}.
The optimized lattice constant in the  $\mathrm{Fe}$@${\mathrm C_6}{\mathrm N_6}$ is found to be $14.20$ \AA\textit{}.

\section{Results and Discussions}
\begin{figure}
{\centering 
\includegraphics[width=6.7in,height=3.0in]{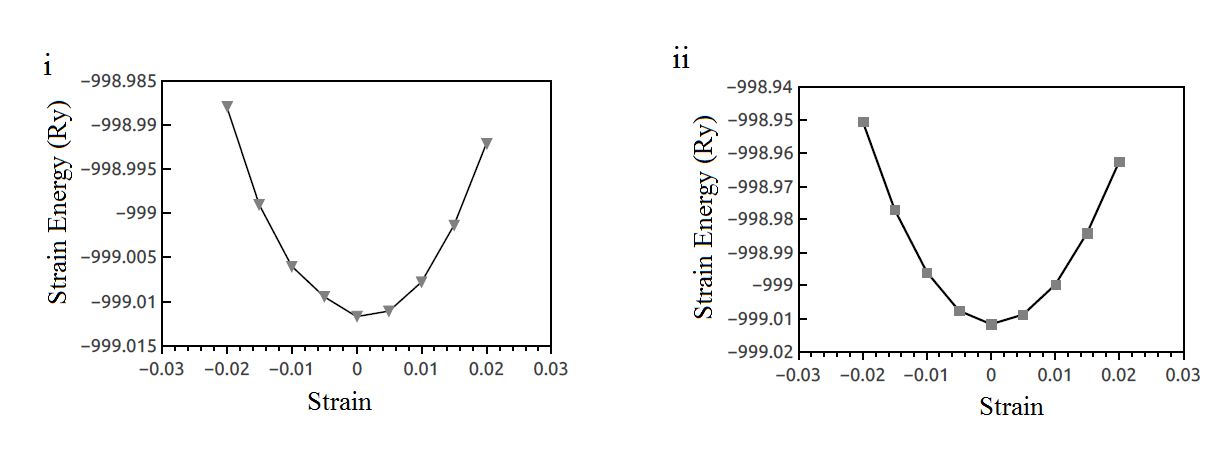} \par}
{\centering  \includegraphics[width=3.0in,height=2.9in]{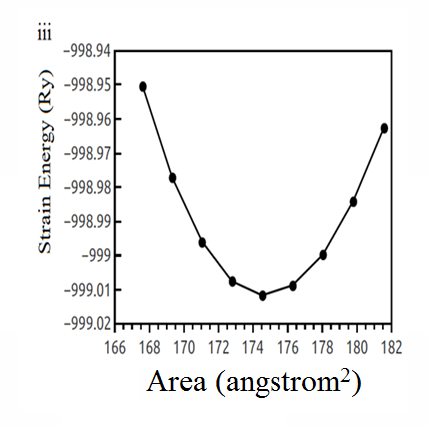} \par}
\caption{Variation of\textbf{ }strain energy (Ry) versus (i) bi-axial tensile strain (ii) uni-axial tensile
strain and (iii) area of the $\mathrm{Fe}$@${\mathrm C_6}{\mathrm N_6}$ system for elastic constant calculation.}
\label{fig2}
\end{figure}

We used the expressions in Eq.~\eqref{eq1} to compute the mechanical properties, such as in-plane stiffness $Y$ (Young modulus), and Poisson's ratio $v$:
\begin{eqnarray}
\label{eq1}
Y&=&m_{11}\left(1-v^2\right) \nonumber \\
v&=&\frac{m_{12}}{m_{11}}.
\end{eqnarray}
The variables $m_{11}$,  $m_{12}$, (known as elastic constants) can be deduced from Eq.~\eqref{eq2}.
\begin{eqnarray}\label{eq2}
m_{11}&=&\frac 1{A_0}\left.\left(\frac{{\partial}^2E}{{\partial}s^2}\right)\right|_{s=0} \ \ \mathrm{(uni-axial)}
\nonumber \\
2\left(m_{11}+m_{12}\right)&=&\frac 1{A_0}\left.\left(\frac{{\partial}^2E}{{\partial}s^2}\right)\right|_{s=0} \ \
\mathrm{(bi-axial)}
\end{eqnarray}
where $A_0$, $E$, and s are equilibrium unit-cell area, strain energy and applied deformation. At each $0.005$
applied deformation (within the harmonic region for both bi- and uni-axial tensile strains, see Fig.~\ref{fig2}) in the 
${xy}${}-plane, the structure is fully relaxed. The calculated in-plane stiffness is 1193.2 $\mathrm{GPa}{\cdot}\text{\AA}$ (= $119.32$ N/m), lower than the in-plane stiffness of pure heptazine sheets \cite{Abdullahi:SSC2016}.
The Poisson's ratio $0.18$ is in the same order as that of graphene sheet \cite{Cadelano:PRB2010}.
The bulk modulus is determined from the product of equilibrium area and the second gradient of deformation energy with respect to area of the 
$\mathrm{Fe}$@${\mathrm C_6}{\mathrm N_6}$ system, which is written as
\begin{equation}
\label{eq3}
G=A \times \left.\left(\frac{{\partial}^2E}{{\partial}A^2}\right)\right|_{A_m}
\end{equation}
The variables in the Eq.~\eqref{eq3} are defined as follows: $A$, $U$ and $A_m$ represent the area of the 
$\mathrm{Fe}$@${\mathrm C_6}{\mathrm N_6}$ sheet, the bi-axial deformation energy and the equilibrium area of  $\mathrm{Fe}$@${\mathrm C_6}{\mathrm N_6}$ respectively. The calculated bulk modulus 86.49 N/m is less than the value for Mn-CN1 system reported in \cite{Abdullahi:CAP2016}.
The difference in the calculated elastic constants in comparisons with previously related structures can be related to the weakening in the bonding networks of the hexagonal rings in the structures. In Table~\ref{table2} we show the values of binding energy $E_b$ and the structural parameters $h$ and $d$ obtained for
different deformations in the range $\pm$ 2\%. By applying strain, the height $h$ (difference between the height of Fe
and the average height of all atoms in the C\textsubscript{6}N\textsubscript{6} sheet) value of the Fe in the relaxed $\mathrm{Fe}$@${\mathrm C_6}{\mathrm N_6}$ sheet do not change. Hence, approximately zero $h$ confirms the planar structure of the 
$\mathrm{Fe}$@${\mathrm C_6}{\mathrm N_6}$ sheet. $E_b$ is determined from Eq.~\eqref{eq4} which is given by:
\begin{equation}
\label{eq4}
E_b=\left(E_{{\mathrm C_6}{\mathrm N_6}}+E_{\mathrm{Fe}}\right)-E_t
\end{equation}
where $E_t$, $E_{{\mathrm C_6}{\mathrm N_6}}$ and  $E_{\mathrm{Fe}}$ denote the total energy of   $\mathrm{Fe}$@${\mathrm C_6}{\mathrm N_6}$ system, the
energy of bare C\textsubscript{6}N\textsubscript{6} sheet, and the total energy of an isolated Fe atom. $E_b$
for unstrained $\mathrm{Fe}$@${\mathrm C_6}{\mathrm N_6}$ system is equal to 4.73 $\mathrm{eV}$. Our computed $E_b$ value is consistent with the previous work \cite{Srinivasu:PCCP2016}.
It can also be seen that the $\mathrm{Fe}$@${\mathrm C_6}{\mathrm N_6}$ structure is more stable for positive binding energy and attained metastable state at a maximum 6\% tensile strain. It is clearly shown in Table~\ref{table2} that
binding energy at a metastable state which is 6\% tensile strain is found to be negative. The decrease in binding
energy as a function of tensile strain can be explained as follows. Based on the definition of strain energy $E$
(i.e., product of square of strain and the elastic constants) under harmonic deformations, an increasing $E$ value stands
for energetically favorable elastic moduli. Thus, for an increasing negative strain energy $E$ of the 
$\mathrm{Fe}$@${\mathrm C_6}{\mathrm N_6}$ system (see Table~\ref{table2}), the binding energy according to Eq.~\eqref{eq4} should correspondingly decrease. Table~\ref{table2} illustrates the uniform increase in $d_{\mathrm{Fe-N}}$ and $d$ bond lengths as a result of distortion in the  $\mathrm{Fe}$@${\mathrm C_6}{\mathrm N_6}$ structure caused by the tensile deformations.

It is known that pure C\textsubscript{6}N\textsubscript{6} sheet is a non-magnetic semiconductor, thus the induced
magnetism in the $\mathrm{Fe}$@${\mathrm C_6}{\mathrm N_6}$ system is mainly from Fe atom (see $M_{\mathrm{Fe}}$ in Table~\ref{table2}). For unstrained $\mathrm{Fe}$@${\mathrm C_6}{\mathrm N_6}$ system, the estimated magnetic moment per unit cell is 3.74 $\mu_{\mathrm B}$, which is consistent
with the recent report \cite{Srinivasu:PCCP2016}.
The magnetic moments are also evaluated for different applied tensile strains. It can be seen
in Table~\ref{table2} that the magnetic moment is less sensitive to the applied strain. This shows that the interaction between the Fe atom and the surrounding atoms in the porous site did not reduce the number of unpaired electrons in the d orbital of Fe atom. According to the results from Lowdin's charge analysis \cite{Lowdin:JCP1950}, the charge transfer $Q$ into C\textsubscript{6}N\textsubscript{6} sheet is contributed mainly by s, p orbitals of Fe atom. There is also a mixed bonding nature (ionic and covalent bonding) which is depicted in the charge-density difference plot (see Fig.~\ref{fig1}(c)) by charge depletion between the atoms and localisation around the most elecronegative N atoms.

\begin{figure}
{\centering  \includegraphics[width=6.4898in,height=2.5937in]{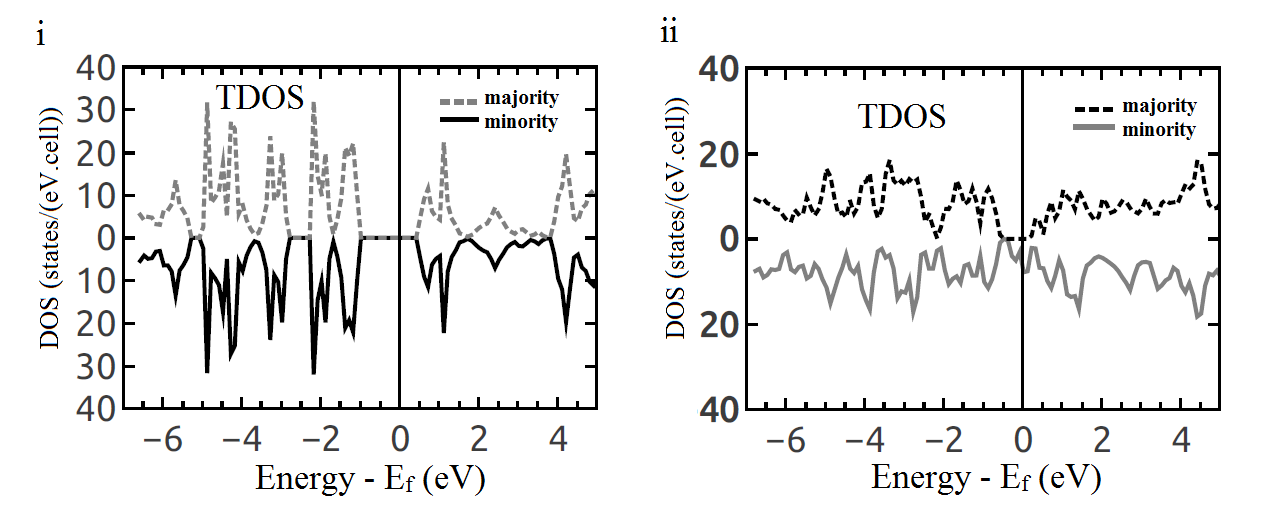} \par}
{\centering  \includegraphics[width=3.2709in,height=2.8in]{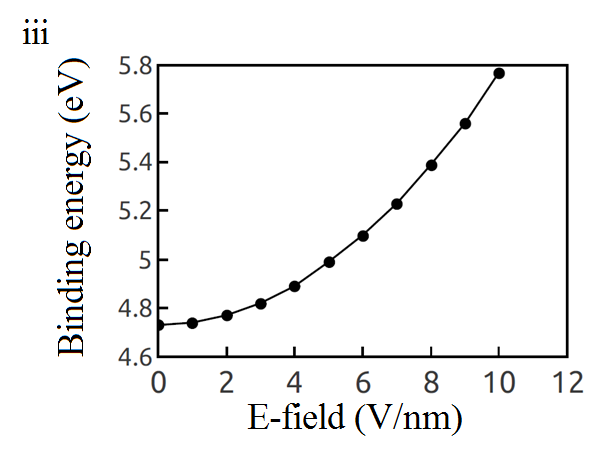} \par}
\caption{
Spin-polarized total density of state (TDOS) for (i) pure s-triazine sheet (ii) $\mathrm{Fe}$@${\mathrm C_6}{\mathrm N_6}$ under applied electric field. (ii) Variation of\textbf{ }binding energy versus applied electric field strength for  $\mathrm{Fe}$@${\mathrm C_6}{\mathrm N_6}$ system. 
}
\label{fig3}
\end{figure}

\begin{figure}
\begin{center}
(a)
\includegraphics[width=2.3in,height=1.8in]{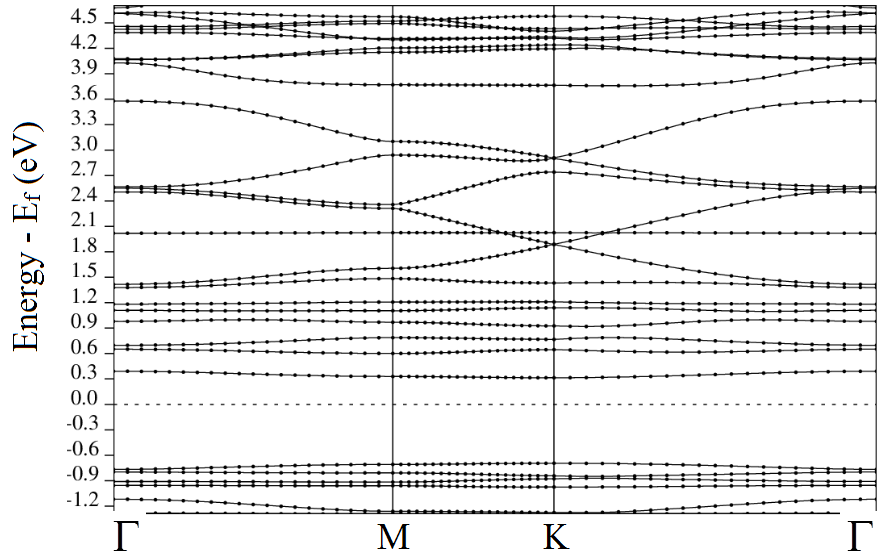}
(b)
\includegraphics[width=2.3in,height=1.8in]{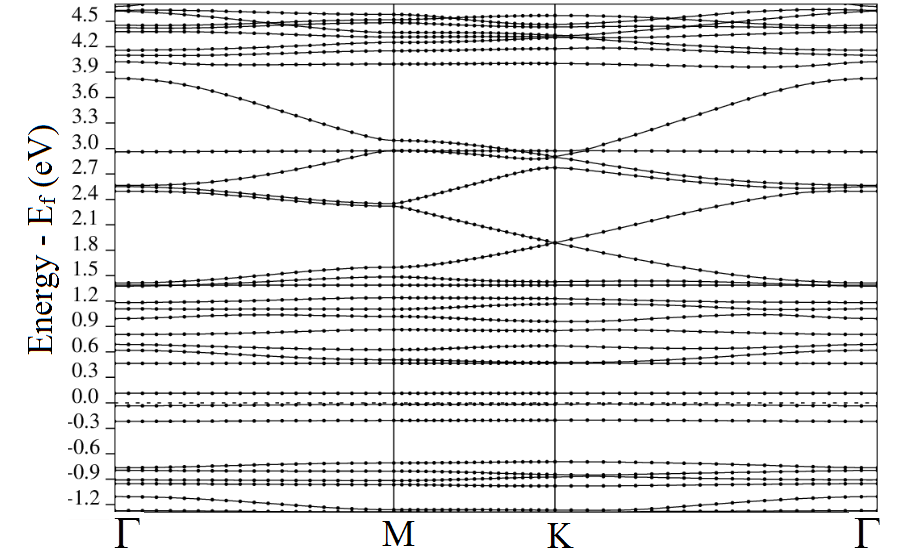} \\
\bigskip
\bigskip
(c) 
{\centering
\includegraphics[width=2.3in,height=1.9in]{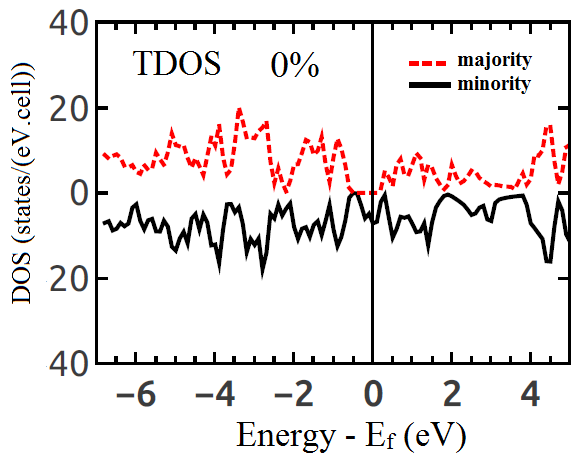}} 
(d)
{\centering
\includegraphics[width=2.3in,height=1.9in]{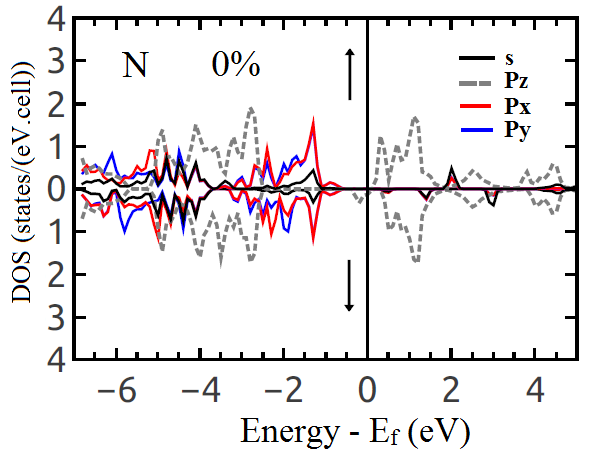} } \\
\bigskip
\bigskip
(e) 
{\centering
\includegraphics[width=2.3in,height=1.9in]{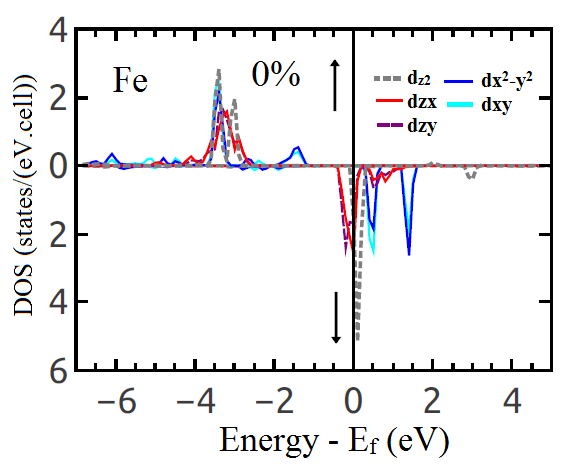}} 
(f)
{\centering
\includegraphics[width=2.3in,height=1.9in]{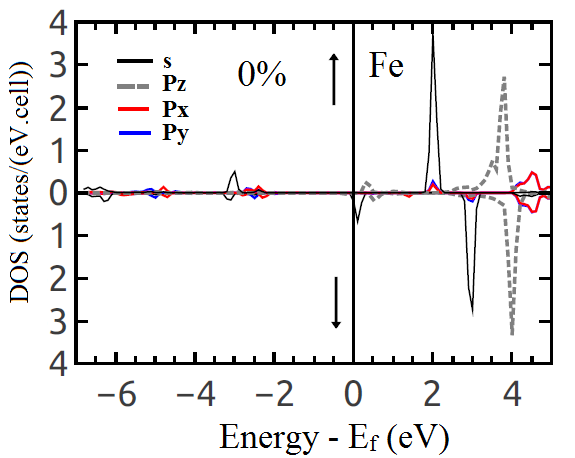}} 
\caption{Spin-polarized electronic band structure (a) majority (b) minority spin states for unstrained
(\textit{s} = 0) $\mathrm{Fe}$@${\mathrm C_6}{\mathrm N_6}$ system. Spin-polarized TDOS and projected density of state (PDOS) for
strain-free (c)-(f).}
\label{fig4}
\end{center}
\end{figure}

To analyze the modulation on the electronic properties of the Fe-embedded s-triazine system, we plotted the electronic
band structure and corresponding total and projected density of states. As can be clearly seen in Fig.~\ref{fig3}(i), pure 
${\mathrm C_6}{\mathrm N_6}$ is nonmagnetic semiconductor. If Fe atom is embedded in the porous site of the ${\mathrm C_6}{\mathrm N_6}$ sheet, the electronic and magnetic properties are modulated. Fig.~\ref{fig4}(a)-(f) show spin-polarised band structure, total density of states (TDOS) and corresponding projected density of state (PDOS) of $\mathrm{Fe}$@${\mathrm C_6}{\mathrm N_6}$ unstrained system. The figures portray half metallic electronic character of the $\mathrm{Fe}$@${\mathrm C_6}{\mathrm N_6}$ system. We can infer that the half-metallic character in $\mathrm{Fe}$@${\mathrm C_6}{\mathrm N_6}$ system is a result of electron transfer from Fe sub-orbital induced by ionic interaction between the Fe and the ${\mathrm C_6}{\mathrm N_6}$ sheet. The minority band structure in Fig.~\ref{fig4}(a) confirms the half-metallic character
which shows flat energy spectrum crossing over the Fermi level at conduction band minimum. From the PDOS plots, we can
observe a dominant features of Fe atom s-, d- sub-orbitals as well as small contributions by $\mathrm{p}_z$- of the
surrounding six N atom in the Fermi level. At approximately -1.5 eV the bonding orbitals are contributed mainly by 
$\mathrm{sp}$-like orbitals of the six N atoms, and the $\mathrm{d}_{{xy}},\mathrm{d}_{x^2-y^2}$ orbitals of Fe atom in the majority spin state. Around -3.5 eV there is a mixed hybridization which is dominated by d-orbitals of the Fe
atom in the majority spin state and s-, $\mathrm{p}_z$-like orbitals of the surrounding N atoms in both majority and
minority spin state.

When subjected to variations in external environment, i.e. symmetric tensile deformation (up to 6\% bi-axial tensile strain, see Table~\ref{table2}) and application of electric field (up to a maximum of 10 V/nm, see Fig~\ref{fig3}(iii)), the half-metallic electronic character and magnetic moment of  $\mathrm{Fe}$@${\mathrm C_6}{\mathrm N_6}$ system  are preserved. On the other hand, binding of Fe atom in $\mathrm{Fe}$@${\mathrm C_6}{\mathrm N_6}$ is enhanced as
larger electric field strength is applied. The resulting higher energy of $\mathrm{Fe}$@${\mathrm C_6}{\mathrm N_6}$ system which corresponds to an increase $E_b$ can be related to repulsive effect between the  $\mathrm{p}_x, \mathrm{p}_y$ orbitals of N atom and the $\mathrm{d}_{{xy}}$, $\mathrm{d}_{x^2-y^2}$ orbitals of Fe atom. This repulsion moves $\mathrm{p}_z$ orbital towards higher energy and hence the system energy increases.

\section{Adsorption of atoms and molecules on Fe embedded striazine sheet}

\begin{figure}
\begin{center}
(a)
\includegraphics[width=2.3in,height=1.9in]{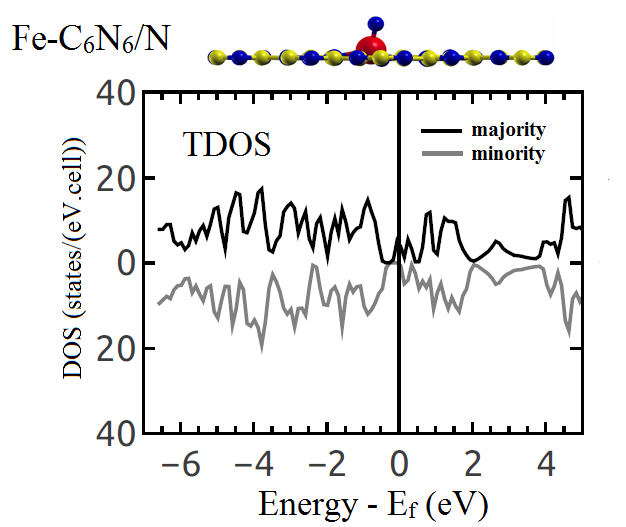}
(b)
\includegraphics[width=2.3in,height=1.9in]{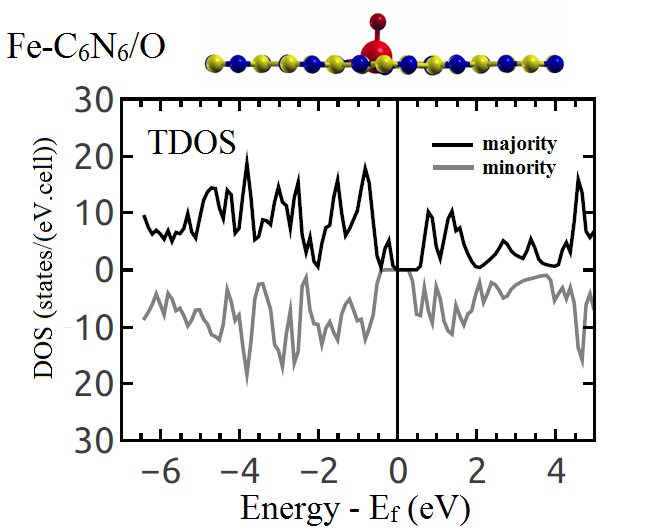} \\
\bigskip
(c)
\includegraphics[width=2.3in,height=1.9in]{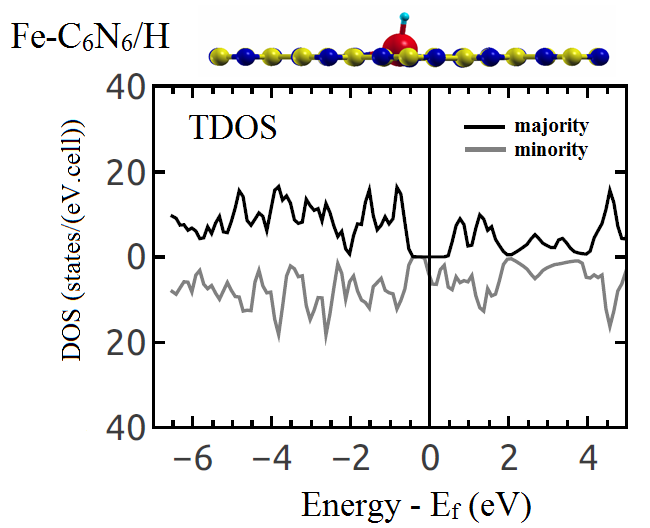}
(d)
\includegraphics[width=2.3in,height=1.9in]{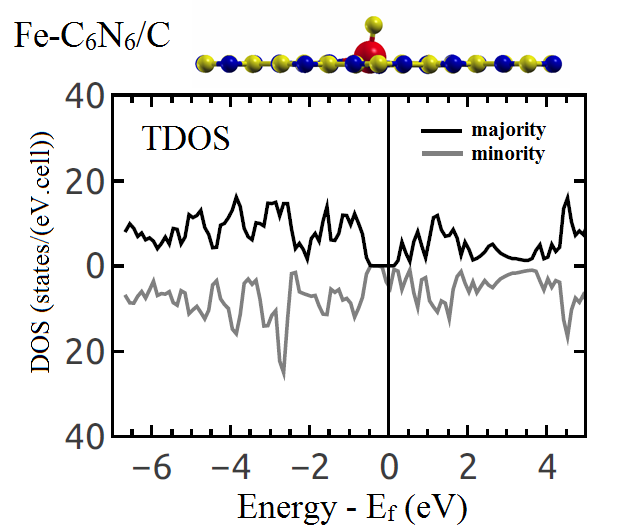}
\caption{Spin-polarized TDOS and side view for $\mathrm{Fe}$@${\mathrm C_6}{\mathrm N_6}$ with an adsorbed (a) C (b) N (c) O and (d) H atoms systems respectively.}
\label{fig5}
\end{center}
\end{figure}

\begin{figure}
\begin{center}
(a)
\includegraphics[width=2.3in,height=1.9in]{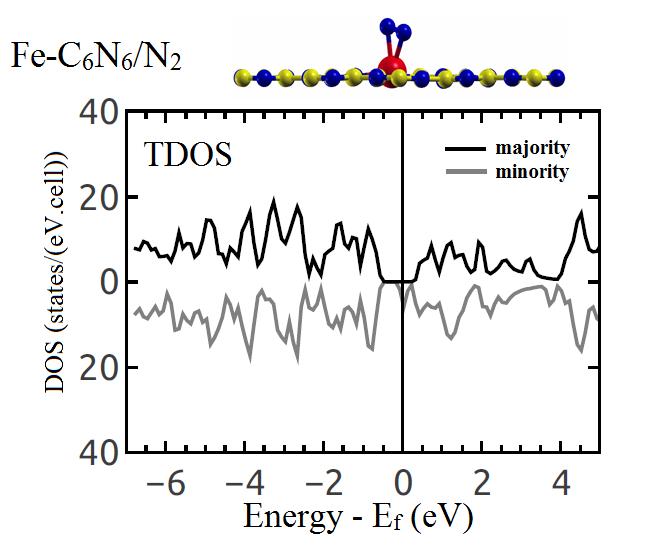}
(b)
\includegraphics[width=2.3in,height=1.9in]{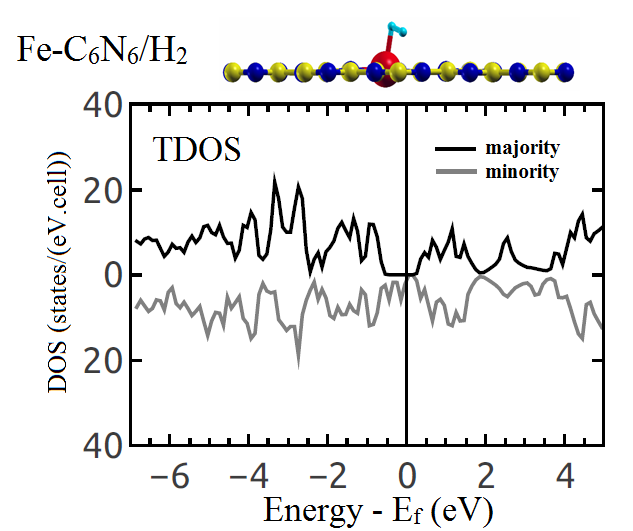} \\
\bigskip
\bigskip
(c)
\includegraphics[width=2.3in,height=1.9in]{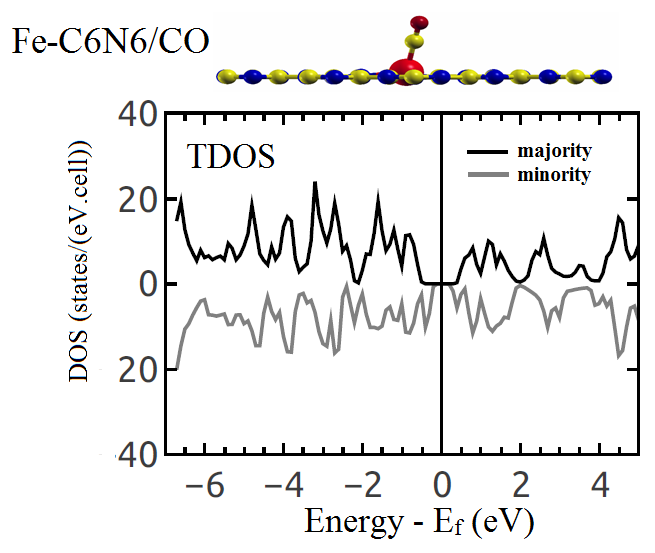} 
(d)
\includegraphics[width=2.3in,height=1.9in]{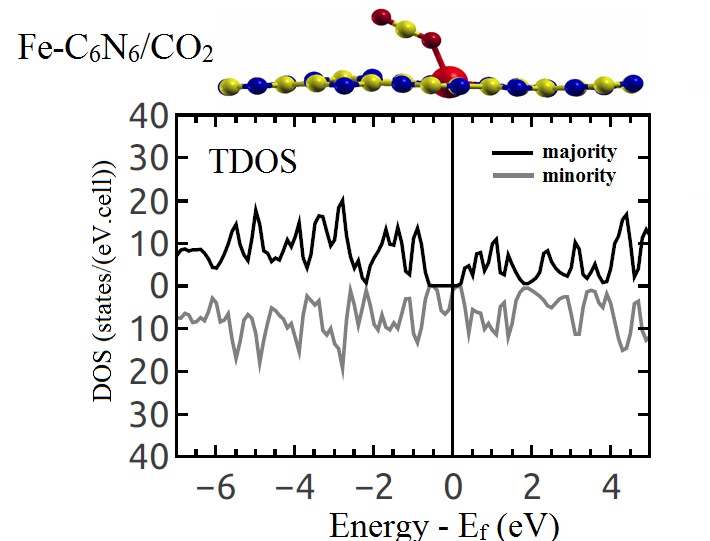} \\
\bigskip
\bigskip
(e)
\includegraphics[width=2.3in,height=1.9in]{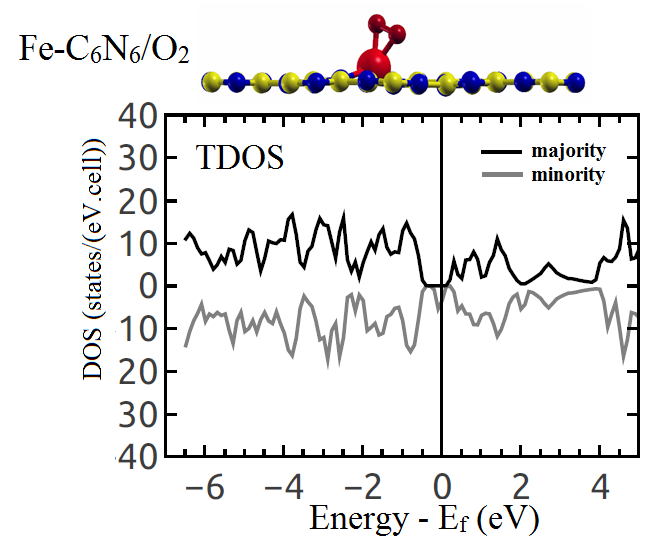}
(f)
\includegraphics[width=2.3in,height=1.9in]{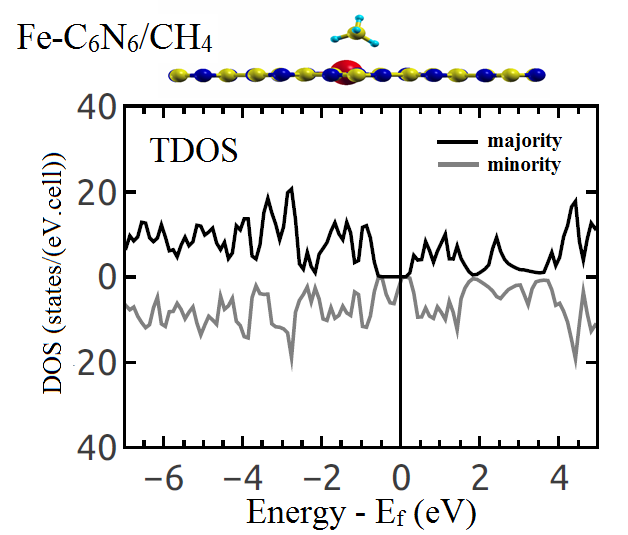}
\caption{Spin-polarized TDOS and side view for  $\mathrm{Fe}$@${\mathrm C_6}{\mathrm N_6}$ with an adsorbed (a) ${\mathrm N_2}$ (b) ${\mathrm H_2}$
(c) CO (d) $\mathrm{CO}_2$ (e) ${\mathrm O_2}$ and (f) $\mathrm{CH}_4$ molecules systems respectively.}
\label{fig6}
\end{center}
\end{figure}

The optimized stable geometries of Fe-embedded s-triazine with adsorbed atoms (C,N,O,H) and molecules
($\mathrm{CH}_4,{\mathrm N_2},{\mathrm O_2},{\mathrm H_2},\mathrm{CO},\mathrm{CO}_2$) systems are displayed in Fig.~\ref{fig5} and \ref{fig6}. All atoms in the system
are allowed to move freely without any constraint during the structural relaxation. We also listed the geometric
parameters of the stable systems in Table~\ref{table3}. For the adsorbed atoms, the parameter  $h_{\mathrm{Fe}-X}$ (where $X$
represents atoms/molecules) are within chemisorption bonding height. These calculated bond lengths are indications of
interaction between the embedded Fe atom and the adsorbed atoms. Nitrogen atom being the most electronegative has the
shortest adsorption height, hence the magnetic moment per unit cell is also the lowest due to the strong interaction. The estimated heights of  $h_{\mathrm{Fe}-X}$ are in agreement with the recent work \cite{Ma:C2015}.
To determine the
stability of the  $\mathrm{Fe}$@${\mathrm C_6}{\mathrm N_6}$ with adsorbed atoms/molecules systems, we calculate the adsorption energy which is expressed as
\begin{equation}
E_{\mathrm{ads}}=\left(E_{\mathrm{Fe}-{\mathrm C_6}{\mathrm N_6}}+E_x\right)-E_T,
\end{equation}
where $E_T$, $E_{\mathrm{Fe}-{\mathrm C_6}{\mathrm N_6}}$ and $E_x$ denotes the total energy of $\mathrm{Fe}$@${\mathrm C_6}{\mathrm N_6}$ with adsorbates, the energy of $\mathrm{Fe}$@${\mathrm C_6}{\mathrm N_6}$ system, and the total energy of an isolated atom or molecules respectively. Positive adsorption energy is an indication of a stable structure. The adsorption energies are listed in
Table~\ref{table3}. The calculated results guarantee the chemical adsorption of all the systems.

As can be seen in Table~\ref{table3}, the adsorption of atoms on $\mathrm{Fe}$@${\mathrm C_6}{\mathrm N_6}$ sheet modulates the total magnetic moment per unit cell. The total magnetic moment in  $\mathrm{Fe}$@${\mathrm C_6}{\mathrm N_6}$ with adsorbed O and H atoms systems increased
whereas we find drastic magnetic moment reduction in   $\mathrm{Fe}$@${\mathrm C_6}{\mathrm N_6}$ with adsorbed N and C. The modulation is related to the electron transfer between the Fe and the surrounding atoms. As illustrated in Table~\ref{table3}, total magnetic moment is high in cases with large charge transfer from Fe into the surrounding atoms as well as the adsorbed atoms (H and O atoms). This shows that interaction between Fe and O, H relatively preserves the number of unpaired electrons in the d orbital of Fe. Thus, the atomic magnetic moment of Fe is maintained in those systems. In contrast, $\mathrm{Fe}$@${\mathrm C_6}{\mathrm N_6}$ with adsorbed N or C atoms produce low spin configurations. TDOS of $\mathrm{Fe}$@${\mathrm C_6}{\mathrm N_6}$ with adsorbed atoms are depicted in Fig.~\ref{fig5}. The TDOS figures for $\mathrm{Fe}$@${\mathrm C_6}{\mathrm N_6}$ with
adsorbed H and O show semiconductor electronic character whereas $\mathrm{Fe}$@${\mathrm C_6}{\mathrm N_6}$ with adsorbed C and N maintain half-metallic character. The modulation of half-metallic character in $\mathrm{Fe}$@${\mathrm C_6}{\mathrm N_6}$ into semiconductor when adsorbed by H and O is caused by the shift in impurity state towards higher energy.

Next we relaxed the structures of $\mathrm{Fe}$@${\mathrm C_6}{\mathrm N_6}$ with six adsorbed molecules ($\mathrm{CH}_4$,  ${\mathrm N_2}$, ${\mathrm O_2}$, ${\mathrm H_2}$, $\mathrm{CO}$, $\mathrm{CO}_2$). The stable adsorption height \textit{h}\textsubscript{Fe-$X$}, the corresponding adsorption energies and the bond lengths $d_{\mathrm{avg}-X}$ of an isolated as well as adsorbed molecules are listed in Table~\ref{table3}. It can be seen that the adsorption heights vary slightly for different molecules. The
smaller value of  $h_{\mathrm{Fe}-X}$ for adsorbed  $\mathrm{CO}$,  ${\mathrm O_2}$ and ${\mathrm H_2}$ is an indication of favourable chemical bonding compared to other adsorbed molecules. Correspondingly, their $d_{\mathrm{avg}-X}$ increases after chemisorption. Interestingly, the adsorption energies for $\mathrm{CO}$ and ${\mathrm O_2}$ are almost the same. Our results are in disagreement with the recent report \cite{Ma:C2015}.
It was reported that ${\mathrm O_2}$ would favourably chemisorbed on the $\mathrm{Fe}$@${\mathrm C_6}{\mathrm N_6}$ surface as compared to $\mathrm{CO}$ when the two gases are passed into the surface at constant
pressure. We used DFT method in our ground state computations, whereas the work in Ref.~\cite{Ma:C2015}
employed DFT+$U$ in the similar adsorption environment (different graphitic CN allotrope). Despite such a difference in computational methodology, we do not expect a large discrepancy in the computed adsorption energy. As it turns out, our result is in the same order as that of Ref.~\cite{Ma:C2015} 
since the $h_{\mathrm{Fe}-X}$ for $\mathrm{CO}$ and ${\mathrm O_2}$ are almost the same. It indicates strong interaction between $\mathrm{CO}$, ${\mathrm O_2}$ molecules and Fe atom due to hybridization. The computed adsorption energy for $\mathrm{CO}_2$ is within the value suggested by Deng \textit{et al.} \cite{Deng:S2013}.

As illustrated in Table~\ref{table3}, high magnetic moment per unit cell are obtained for different $\mathrm{Fe}$@${\mathrm C_6}{\mathrm N_6}$ with adsorbed molecules systems. As can be seen under $M\textsubscript{atom}$, the contributions of the magnetic moment in the systems mainly originates from Fe atom. This shows that the high spin configuration of 3d electrons of the Fe atom are maintained. Except for $\mathrm{CO}$ and N\textsubscript{2} which couple antiferromagnetically with the Fe atoms, the rest of the molecules aligned in the same order with the Fe atom. Hence, the increase or decrease in the number of unpaired electrons in the 3d orbitals of the Fe atom determines the total magnetic moment. Figs.~\ref{fig6} show the geometries and TDOS of $\mathrm{Fe}$@${\mathrm C_6}{\mathrm N_6}$ with adsorbed molecules. It is seen that the half-metallic character of $\mathrm{Fe}$@${\mathrm C_6}{\mathrm N_6}$ system can be tuned into a semiconducting one via adsorption of ${\mathrm H_2}$, ${\mathrm O_2}$ and $\mathrm{CH}_4$ molecules onto its surface.

\section{Conclusion}
In summary, we have investigated the mechanical, geometrical, electronic and magnetic properties of $\mathrm{Fe}$@${\mathrm C_6}{\mathrm N_6}$ system under the influence of external environment based on first-principles calculations. Our findings reveal that the binding energy of $\mathrm{Fe}$@${\mathrm C_6}{\mathrm N_6}$ can be modulated by an applied tensile strain and perpendicular electric field. The non-magnetic semiconducting property of bare C\textsubscript{6}N\textsubscript{6} is modulated upon embedding of Fe atom in the porous site of the sheet. It is found that the $\mathrm{Fe}$@${\mathrm C_6}{\mathrm N_6}$ system exhibits half-metallic electronic character with magnetic moment which is in the order of that for an isolated Fe atom. Additionally, the electronic and magnetic properties of the $\mathrm{Fe}$@${\mathrm C_6}{\mathrm N_6}$ systems are preserved under a maximum value of 10 V/nm in electric field strength and 6\% tensile strain. 

Interestingly, we find that the half-metallic electronic character of $\mathrm{Fe}$@${\mathrm C_6}{\mathrm N_6}$ system can be tuned into a semiconductor via adsorption of atoms and molecules into the $\mathrm{Fe}$@${\mathrm C_6}{\mathrm N_6}$ system. The appearance of semiconducting character is a result of shift in impurity state towards higher energy when
the atoms or molecules are adsorbed on $\mathrm{Fe}$@${\mathrm C_6}{\mathrm N_6}$ surface. The magnetic moment of $\mathrm{Fe}$@${\mathrm C_6}{\mathrm N_6}$ with adsorbed atoms/molecules is also modified. Our findings may serve as a guide for future applications of $\mathrm{Fe}$@${\mathrm C_6}{\mathrm N_6}$  structures in spintronics devices.

\section*{Acknowledgments}

T. L. Yoon wishes to acknowledge the support of Universiti Sains Malaysia RU grant (No. 1001/PFIZIK/811240). Figures
showing atomic model and 2D charge-density difference plots are generated using the XCRYSDEN program Ref.~\cite{Kokalj:CMS2003}. 
We gladfully acknowledge Dr. Chan Huah Yong from USM School of Computer Science, and Prof. Mohd. Zubir Mat Jafri from USM School of Physics, for providing us computing resources to carry out part of the calculations done in this paper.

\bibliographystyle{elsarticle-num}
\bibliography{refSixpaper}

\begin{table}[p]
\caption{Calculated lattice parameters and total strain energy for Fe atom embedded striazine system.}
\label{table1}
\begin{center}
\tablefirsthead{}
\tablehead{}
\tabletail{}
\tablelasttail{}
\begin{supertabular}{m{0.6136598in}m{0.8906598in}m{0.9643598in}m{0.9684598in}m{0.8677598in}}
\hline
{\centering Strain\par}

\centering (\%) &
{\centering Area\par}

\centering (\AA \textsuperscript{2}) &
{\centering Total energy\par}

{\centering Biaxial\par}

\centering (Ry) &
{\centering Total energy\par}

{\centering Uniaxial\par}

\centering (Ry) &
{\centering Lattice parameter\par}

{\centering Biaxial\par}

\centering\arraybslash (\AA)\\\hline
\centering {}-0.02 &
\centering 167.61 &
\centering {}-998.95042 &
{}-998.98795 &
\centering\arraybslash 13.91/24.10\\
\centering {}-0.015 &
\centering 169.32 &
\centering {}-998.97718 &
{}-998.99904 &
\centering\arraybslash 13.98/24.22 \ \ \\
\centering {}-0.01 &
\centering 171.05 &
\centering {}-998.99598 &
{}-999.00600 &
\centering\arraybslash 14.05/24.34\\
\centering {}-0.005 &
\centering 172.78 &
\centering {}-999.00762 &
{}-999.00946 &
\centering\arraybslash 14.13/24.47\\
\centering 0 &
\centering 174.52 &
\centering {}-999.01171 &
{}-999.01171 &
\centering\arraybslash 14.20/24.59\\
\centering 0.005 &
\centering 176.27 &
\ \ \ \ {}-999.00880 &
{}-999.01108 &
\centering\arraybslash 14.27/24.71\\
\centering 0.01 &
\centering 178.03 &
\centering {}-998.99980 &
{}-999.00779 &
\centering\arraybslash 14.34/24.83\\
\centering 0.015 &
\ \ \ \ \ 179.79  &
\centering {}-998.98403 &
{}-999.00136 &
\centering\arraybslash 14.41/24.96\\
\centering 0.02 &
\centering 181.57 &
\centering {}-998.96248 &
{}-998.99209 &
\centering\arraybslash 14.48/25.08\\\hline
\end{supertabular}
\end{center}
\end{table}

\begin{table}[p]
\caption{The calculated binding energies $E_b$, the average bond length between Fe atom and N\textsubscript{edge} atoms  $d_{\mathrm{Fe}-N}$, average bond length connecting the s-triazine $d$, and Fe height $h$\textit{ }(refers to the difference in the \textit{z}{}-coordinate of the Fe atom and the average of the \textit{z}{}-coordinate of all the C and N atoms in the  ${\mathrm C_6}{\mathrm N_6}$ sheet). Charge transfer, magnetic moment per unit cell and per Fe atom,\textcolor{black}{ electronic character of the}   $\mathrm{Fe}$@${\mathrm C_6}{\mathrm N_6}$ system
are denoted by $Q$, $M_{\mathrm{cell}}$, $M_{\mathrm{Fe}}$, $\mathrm{EC}$ respectively. All the systems are half-metallic.}
\label{table2}
\begin{center}
\tablefirsthead{}
\tablehead{}
\tabletail{}
\tablelasttail{}
\begin{supertabular}{m{0.48585984in}m{0.49485984in}m{0.95595986in}m{0.59345984in}m{0.5094598in}m{0.7844598in}m{0.37545985in}m{0.42125985in}m{0.42475984in}}
\hline
\\
\centering Strain &
{\centering  $E_b$ \par}

\centering (eV) &
{\centering  $d_{\mathrm{Fe-N}}$ \par}

\centering (\AA) &
{\centering  $d$\par}
\centering (\AA) &
{\centering  $h$\par}
\centering (\AA) &
{\centering  $Q$\par}
\centering (electrons) &
{\centering  $M_{\mathrm{Fe}}$\par}
\centering (\textit{{\textmu}}\textsubscript{B}) &
{\centering  $M_{\mathrm{cell}}$ \par}

\centering (\textit{{\textmu}}\textsubscript{B}) &
\centering\arraybslash  $\mathrm{EC}$\\\hline
\centering 0\% &
\centering 4.73 &
\centering 2.06 -- 3.40 &
\centering 1.49 &
\centering {}-0.01 &
\centering 0.53 &
\centering 3.61 &
\centering 3.74 &
\centering\arraybslash HM\\
\centering 1\% &
\centering 4.56 &
\centering 2.06 -- 3.49 &
\centering 1.51 &
\centering {}-0.01 &
\centering 0.54 &
\centering 3.61 &
\centering 3.73 &
\centering\arraybslash HM\\
\centering 2\% &
\centering 4.05 &
\centering 2.09 -- 3.56 &
\centering 1.53 &
\centering 0.00 &
\centering 0.56 &
\centering 3.61 &
\centering 3.71 &
\centering\arraybslash HM\\
\centering 3\% &
\centering 3.21 &
\centering 2.09 -- 3.64 &
\centering 1.55 &
\centering 0.00 &
\centering 0.56 &
\centering 3.61 &
\centering 3.70 &
\centering\arraybslash HM\\
\centering 4\% &
\centering 2.06 &
\centering 2.02 -- 3.80 &
\centering 1.58 &
\centering 0.00 &
\centering 0.53 &
\centering 3.57 &
\centering 3.69 &
\centering\arraybslash HM\\
\centering 5\% &
\centering 0.75 &
\centering 2.08 -- 3.82 &
\centering 1.61 &
\centering 0.00 &
\centering 0.54 &
\centering 3.57 &
\centering 3.66 &
\centering\arraybslash HM\\
\centering 6\% &
\centering {}-0.82 &
\centering 2.08 -- 3.91 &
\centering 1.63 &
\centering 0.00 &
\centering 0.56 &
\centering 3.57 &
\centering 3.64 &
\centering\arraybslash HM\\\hline
\end{supertabular}
\end{center}
\end{table}

\begin{table}[p]
\caption{$E_{\mathrm{ads}}$ denotes calculated adsorption energy. $h_{\mathrm{Fe}-X}$ denotes averaged height 
between Fe atom and the adsorbates.  $d_{\mathrm{avg}-X}$ denotes bond length of molecules. $X$ represents adsorbate
species. The values without parenthesis are that for absorbed molecules while that in parenthesis are for isolated
molecules. $Q$ refers to net charge transfer among the adsorbates and the $\mathrm{Fe}$@${\mathrm C_6}{\mathrm N_6}$ system. The values without parenthesis are charge transfer from Fe atom into the sheet or adsorbates. These values are all positive as electron from Fe atom always get transferred into the surrounding. The $Q$ values in parenthesis are charge transfer into the Fe-striazine sheet from adsorbates. Positive values mean electron is transferred into the
surroundings ($\mathrm{Fe}$@${\mathrm C_6}{\mathrm N_6}$ system) from adsorbates, and vice versa. $M_{\mathrm{cell}}$ refers to magnetic moment per unit cell. $M_{\mathrm{atom}}$ refers to magnetic moment of Fe atom or adsorbates. The values without parenthesis are that for Fe atom, while that in parenthesis are that for the adsorbates. $\mathrm{EC}$ refers to the electronic character of the  $\mathrm{Fe}$@${\mathrm C_6}{\mathrm N_6}$ with adsorbates. In the present case,  $\mathrm{EC}$ can be either half metallic (HM) or semiconducting (SC).} 
\label{table3}
\begin{center}
\tablefirsthead{}
\tablehead{}
\tabletail{}
\tablelasttail{}
\begin{supertabular}{m{0.5615598in}m{0.41845986in}m{0.5448598in}m{0.6080598in}m{1.4219599in}m{0.39005986in}m{0.82335985in}m{0.35805985in}}
\hline
\\
\centering System &
{\centering  $E_{\mathrm{ads}}$\par}
\centering (eV) &
{\centering  $h_{\mathrm{Fe}-X}$\par}
\centering (\AA) &
{\centering  $d_{\mathrm{avg}-X}$\par}
\centering (\AA) &
{\centering  $Q$\par}
\centering (electrons) &
{\centering  $M_{\mathrm{cell}}$\par}
\centering (\textit{{\textmu}}\textsubscript{B}) &
{\centering  $M_{\mathrm{atom}}$\par}
\centering (\textit{{\textmu}}\textsubscript{B}) &
\centering\arraybslash  $\mathrm{EC}$ \\ 
\\
\hline
\centering C &
\centering 4.48 &
\centering 1.56 &
\centering {}- &
{\centering 0.13\par}

\centering (0.19) &
\centering 1.68 &
{\centering 2.17\par}

\centering (-0.71) &
\centering\arraybslash \ HM\\
\centering N &
\centering 3.00 &
\centering 1.52 &
\centering {}- &
{\centering 0.24\par}

\centering (0.02) &
\centering 1.56 &
{\centering 1.53\par}

\centering (-0.12) &
\ \ \ HM\\
\centering O &
\centering 4.11 &
\centering 1.65 &
\centering {}- &
{\centering 0.54\par}

\centering (-0.34) &
\centering 4.57 &
{\centering 3.45\par}

\centering (0.78) &
\centering\arraybslash SC\\
\centering H &
\centering 1.90 &
\centering 1.60 &
\centering {}- &
{\centering 0.46\par}

\centering (-0.12) &
\centering 4.31 &
{\centering 3.93\par}

\centering (0.11) &
\centering\arraybslash SC\\
\centering CO &
\centering 1.30 &
\centering 1.89 &
{\centering 1.14\par}

\centering (1.16) &
{\centering 0.29\par}

\centering (C: 0.33, O: 0.09) &
\centering 3.53 &
{\centering 3.48\par}

\centering (-0.13) &
\centering\arraybslash SC\\
\centering CO\textsubscript{2} &
\centering 0.44 &
\centering 2.16 &
{\centering 1.18\par}

\centering (1.18) &
{\centering 0.50\par}

\centering (C: 0.79, O: -0.41) &
\centering 3.78 &
{\centering 3.64\par}

\centering (0.24) &
\centering\arraybslash HM\\
\centering O\textsubscript{2} &
\centering 1.32 &
\centering 1.82 &
{\centering 1.23\par}

\centering (1.40) &
{\centering 0.52\par}

\centering (-0.23) &
\centering 3.70 &
{\centering 3.12\par}

\centering (0.49) &
\centering\arraybslash HM\\
\centering N\textsubscript{2} &
\centering 0.15 &
\centering 2.12 &
{\centering 1.11\par}

\centering (1.14) &
{\centering 0.36\par}

\centering (0.2) &
\centering 3.55 &
{\centering 3.51\par}

\centering (-0.15) &
\centering\arraybslash HM\\
\centering H\textsubscript{2} &
\centering 0.31 &
\centering 1.87 &
{\centering 0.75\par}

\centering (0.79) &
{\centering 0.42\par}

\centering (0.21) &
\centering 3.68 &
{\centering 3.53\par}

\centering (0.00) &
\centering\arraybslash SC\\
\centering CH\textsubscript{4} &
\centering 0.18 &
\centering 2.16 &
{\centering 1.10\par}

\centering (1.10) &
{\centering 0.51\par}

\centering (C: -0.64, H: 0.90) &
\centering 3.76 &
{\centering 3.62\par}

\centering (0.02) &
\ \ SC \\\hline
\end{supertabular}
\end{center}
\end{table}

\end{document}